\documentclass[12pt]{article}

\usepackage[latin1]{inputenc}
\usepackage{hyperref}
\usepackage{graphicx,natbib,amsmath,colonequals}
\usepackage{color}
\usepackage{url,multirow}
\usepackage{authblk}
\usepackage[margin=1in]{geometry}
\usepackage{amsthm}
\newtheorem{proposition}{Proposition}
\usepackage[font={small}]{caption,subcaption}

\title{BADER: Bayesian analysis of differential expression in RNA sequencing data}

\author[1]{Matthias Katzfuss\thanks{Corresponding author. Email: \href{mailto:katzfuss@gmail.com}{\nolinkurl{katzfuss@gmail.com} }}}
\author[2]{Andreas Neudecker}
\author[3]{Simon Anders}
\author[4]{Julien Gagneur}
\affil[1]{Texas A\&M University}
\affil[2]{Universit\"{a}t Jena}
\affil[3]{European Molecular Biology Laboratory, Heidelberg}
\affil[4]{Gene Center, Ludwig-Maximilians-Universit\"{a}t M\"{u}nchen}

\date{}

\begin{document}

\maketitle

\begin{abstract}
Identifying differentially expressed genes from RNA sequencing data remains a
challenging task because of the considerable uncertainties in parameter
estimation and the small sample sizes in typical applications.
Here we introduce Bayesian Analysis of
Differential Expression in RNA-sequencing data (BADER). 
Due to our choice of data and prior distributions, full posterior inference for BADER can be carried out
efficiently. The method appropriately takes uncertainty in gene variance into
account, leading to higher power than existing methods in detecting differentially expressed
genes. Moreover, we show that the posterior samples can be naturally
integrated into downstream gene set enrichment analyses, with excellent performance in detecting enriched sets.
An open-source \texttt{R} package (BADER) that provides a user-friendly interface to a \texttt{C++} back-end is available on Bioconductor.
\end{abstract}


\section{Introduction}

Transcription profiling by deep sequencing (RNA-seq) has become the technology
of choice for quantifying gene expression, because of its high throughput, its
large dynamic range, and the possibility to identify novel transcripts.
However, noise in RNA-seq experiments is not yet well understood. Hence,
appropriate statistical models to quantify gene expression and assess
statistical significance of differential expression (DE) are needed and are the
object of intense research \citep[e.g.,][]{Garber2011}.

Due to biological variation, gene-level RNA-seq count data are overdispersed
relative to the Poisson distribution. Most methods assume that the
read counts come from negative binomial distributions and differ in their
treatment of the (over-)dispersion parameters.
The edgeR method \citep{Robinson2008,robinson2010edger} estimates 
dispersion parameters of individual genes robustly by shrinking them toward a
genome-wide parametric function relating dispersion to mean, whereas
DEseq \citep{Anders2010} models this relation nonparametrically.
\citet{Lund2012} noticed the importance of taking uncertainty
in the estimation of the dispersion parameter into account and were able to
integrate it into their statistical test. More recently, \citet{Wu2013}
observed systematic over- and under-shrinkage of the gene-specific
dispersion parameters in edgeR and DEseq, respectively. To circumvent
this issue, \citet{Wu2013} stayed within a frequentist paradigm but placed a prior
on dispersion parameters and thus were able to obtain better DE detection in their analyses.

Moreover, calling DE genes is typically the starting
point for further analyses, most prominently gene set enrichment analyses.
However, the power for calling a gene DE typically increases with
its average read counts. Frequentist approaches that do not take this variation in power into account are bound to mischaracterize uncertainty regarding whether sets are enriched.
To cope with the difference in testing power among genes, dedicated gene set
enrichment methods have been investigated \citep{Young2010,Mi2012}.

Altogether, the limitations met by frequentist approaches --- for example, properly quantifying uncertainty for identifying
DE genes and for downstream analyses
--- call for a fully Bayesian treatment. \citet{Hardcastle2010} followed an
empirical Bayesian approach called baySeq. However, the prior parameters are
estimated using quasi-likelihood methods, which ignores uncertainty in the parameter estimation. Two
Bayesian approaches, MMseq \citep{Turro2011}, which focuses on Poisson noise, and
the more recent Bitseq \citep{Glaus2012}, which accounts for overdispersed
data, were developed to tackle the more complicated problem of estimating expression levels
of transcript isoforms originating from the same locus, leading to highly coupled
estimates. In order to address this difficult task, BitSeq required some
shortcuts in the inference procedure. Moreover, the approach does not return explicit
posterior probabilities for genes to be differentially expressed. Instead, the authors advise to use the ranking of the most probable up- and
down-regulated genes.


Here, we propose a fully Bayesian analysis of differential expression in RNA
sequencing data (BADER). Because we perform full posterior inference, our
approach takes account of all uncertainties in a natural way and is theoretically
justified.

BADER allows the dispersion parameters to be gene-specific, yet it is robust even
for datasets with very small sample sizes due to borrowing of information across
genes through a common hyperprior.
We model the log fold change of a gene (\emph{i.e.}, the log ratio of the mean
expression levels between two groups) by a continuous random variable
with a point mass at zero. This provides a probabilistic indicator whether a gene
exhibits true DE, or if the difference in expression is
caused by random variability. It also yields a density estimate for the magnitude
of the log fold change in the case of DE. Using simulated data, we show that
BADER exhibits excellent performance in terms of DE detection.

Finally, we demonstrate how posterior samples of the DE indicators can be integrated into downstream
analyses in the case of gene set enrichment. Our results show that this approach
has more power to detect enriched gene sets than a frequentist counterpart.

\section{Methods}
\subsection{Model Description \label{sec:model}}

Assume that from an RNA-seq experiment we have a set of read counts of the form:
\begin{equation}
\label{data}
\{k_{ij}^T \!: \; i = 1,\dots,n_T ; \, j=1,\dots,m; \, T=A,B \},
\end{equation}
where $i$ denotes the library or sample, $j$ denotes the genomic region, and $T$
denotes the experimental condition (treatment). ``Genomic region'' here could be a gene, an exon, or a set of exon, but henceforth we will refer to it as a gene for simplicity.

In many recent publications \citep[e.g.,][]{Robinson2008,Anders2010}, the
negative-binomial distribution is the overdispersed-Poisson data model of
choice for such data. This distribution can be written as the marginal
distribution of a Poisson random variable whose rate parameter follows a gamma
distribution. Here we assume that the Poisson rate parameter follows a
lognormal distribution, leading to the following two-stage data model (see Appendix \ref{app:analyticcomparison} for more details):
\begin{equation}
\label{lognormalmodel}
\begin{split}
   k_{ij}^T | \lambda_{ij}^T & \stackrel{ind.}{\sim} Poi( s_i^T e^{ \lambda_{ij}^T}); \quad \text{for all} \; i, j, T, \\
   \lambda_{ij}^T | \mu_j^T, \alpha_j^T & \stackrel{ind.}{\sim} N( \mu_j^T, e^{\alpha_j^T}  ); \quad \text{for all} \; i, j, T,
\end{split}
\end{equation}
where $s_i^T$ denotes the \textit{sampling depth} of sample or library $i$ in
group $T$, which we estimate using the median ratio estimator of
\citet{Anders2010}. The data model \eqref{lognormalmodel} was chosen for
computational reasons. While inference from the gamma and lognormal
distribution typically leads to the same conclusions (\citealp{Atkinson1982};
\citealp{McCullagh1989}, pp.\ 286/293), the
latter allows us to obtain many MCMC parameter updates in closed form (see
Section \ref{sec:inference} below), which results in more efficient MCMC
sampling (see Appendix \ref{app:numericalcomparison}).

The log mean rates for the two groups $A$ and $B$ are noted by $\mu_j^A$
and $\mu_j^B$, respectively. For $\mu_j^A$, we
assume a noninformative prior with density $[\mu_j^A] \propto 1$, independently
for $j = 1,\dots,m$. A crucial component of our model is the log-fold-change
parameter, $ \gamma_j \colonequals \mu_j^B - \mu_j^A$, whose prior distribution
is assumed to be a mixture distribution of a point mass at zero --- indicating
no differential expression (DE) for gene j --- and a normal distribution: 
\begin{equation}
\label{foldchange}
	\gamma_j | I_j, \sigma_\gamma^2 \stackrel{ind.}{\sim}
           \begin{cases} 0, & I_j = 0 \\ N(0,\sigma_\gamma^2), & I_j = 1 \end{cases} ; \quad
            j=1,\dots,m,
\end{equation}
where $I_j | \pi \stackrel{iid}{\sim} B(\pi) $, $j=1,\dots,m$, are Bernoulli DE indicators.

The parameters $\alpha_j^T$ in \eqref{lognormalmodel} determine the degree of
overdispersion (for $\alpha_j^T \rightarrow - \infty$, the data model for
$k_{ij}^T$ in \eqref{lognormalmodel} is exactly Poisson). We assume
\begin{equation}
\label{dispersionprior}
\alpha_j^T \stackrel{iid}{\sim} N(\psi_0,\tau^2); \quad j=1,\ldots,m; \, T = A,B.
\end{equation}

For all hyperparameters in \eqref{foldchange} and \eqref{dispersionprior}, we
assume independent noninformative priors: $\pi \sim U(0,1)$, $[\psi_0] \propto
1$, $[\sigma_\gamma^2 ] \propto 1/\sigma_\gamma^2$, and $[\tau^2] \propto
1/\tau^2$. The latter two priors are the Jeffreys prior for the variance
parameter of a normal distribution.

\subsection{Posterior Inference \label{sec:inference}}

Given a set of RNA-seq counts, $\{k_{ij}^T\}$, as in \eqref{data}, the primary interest is in the
posterior distribution of the log-fold-change parameters, $\{ \gamma_j \}$. Once
we have obtained this (joint) distribution, it is trivial to derive quantities of
interest, like the individual posterior DE probabilities, $ p_j \colonequals
P(I_j = 1 | \{k_{ij}^T\})$; $j=1,\dots,m$. A great advantage of Bayesian
inference is that we obtain \emph{joint} posterior distributions, and so we do
not have to account for multiple testing explicitely (see,
\citep[e.g.,][]{Scott2006}. We can also use the results for further downstream analyses
(e.g., for gene set enrichment; see Section \ref{sec:enrichment} below).

Because there is no closed-form expression for the joint posterior of all
unknown variables in the model, we resort to sampling-based inference via a Markov chain
Monte Carlo (MCMC) algorithm. We use a Gibbs sampler \citep{Geman1984}, which
cycles through all unknowns and updates each parameter based on its so-called
full conditional distribution (FCD). For a generic parameter (or set of
parameters) $\theta$, the FCD is defined as the conditional distribution of
$\theta$ given the data and all other variables in the model, and we denote the
FCD as $[\theta | \,\cdot\,]$.

The FCDs of the rate and dispersion parameters are not available in closed form,
\begin{align*}
\textstyle[\lambda_{ij}^T | \, \cdot \, ]\propto\, & \textstyle Poi( k_{ij}^T | s_i^T e^{\lambda_{ij}^T} ) \,N(\lambda_{ij}^T | \mu_j^A + \gamma_j I(T\!=\!B), \exp\{\alpha_j^T\}) , \displaybreak[0]\\
\textstyle[\alpha_j^T | \,\cdot \, ]\propto\, &\textstyle\big( \prod_{i} N(\lambda_{ij}^T | \mu_j^A +  \gamma_j I(T\!=\!B), \exp\{\alpha_j^T\}) \big) N(\alpha_j^T | \psi_0 ,\tau^2),
\end{align*}
and so we use adaptive Metropolis-Hastings steps \citep{Haario2001} to update these parameters for all $i,j,T$.

The updates of the hyperparameters are given by:
\begin{align*}
	\pi |\,\cdot\,\sim\,& \textstyle Beta(1+ \sum_{j} I_j, 1+ \sum_{j} (1-I_j) ), \displaybreak[0]\\
	\sigma_\gamma^2 |\, \cdot\, \sim \,&\textstyle InvGamma(\sum_{j} I_j/2,\sum_{j } \gamma_j^2 I_j /2), \displaybreak[0]\\
	\psi_0 |\, \cdot \,\sim\,& \textstyle N (\sum_{j,T}\alpha_j^T/2m,\tau^2/2m), \\
	\tau^2 |\, \cdot\, \sim\,&\textstyle InvGamma(m, \sum_{j,T}(\alpha_j^T-\psi_0)^2/2).
\end{align*}

Initial tests showed heavy posterior dependence between $I_j$, $\mu_j^A$ and
$\gamma_j$. Fortunately, our choice of the lognormal distribution in
\eqref{lognormalmodel} allows updating these parameters jointly from $[\gamma_j,
\mu_j^A, I_j | \cdot ]$ by integrating both $\mu_j^A$ and $\gamma_j$ out when
updating $I_j$ and by integrating $\gamma_j$ out when updating $\mu_j^A$. Let
$\Omega$ denote the set of all parameters, and define $\overline{\lambda_j^T}
\colonequals  \sum_i \lambda_{ij}^T/n_T$, $v_0 \colonequals \exp\{\alpha_j^B\}$
and $v_1\colonequals \exp\{\alpha_j^B\} + n_T \sigma_\gamma^2$. The updates are
as follows:
\begin{align*}
\textstyle I_j | \Omega \backslash \{\mu_j^A, \gamma_j, I_j\} & \textstyle\stackrel{ind.}{\sim} B\big(\pi N_1/( \pi N_1 + (1-\pi) N_0) \big), \; \text{where} \\
N_l & \textstyle\colonequals N\big(\overline{\lambda_j^A} | \overline{\lambda_j^B}, (\exp\{\alpha_j^A\} + v_l)/n_T\big); \quad l \in \{0,1\}, \displaybreak[0]\\
\textstyle\mu_j^A |  \Omega \backslash \{\mu_j^A, \gamma_j \} &\textstyle\stackrel{ind.}{\sim} 
N\Big(\frac{\overline{\lambda^A_j} v_{I_j} + \overline{\lambda_j^B} \exp\{\alpha_j^A\}}{\exp\{\alpha_j^A\} + v_{I_j}},\frac{\exp\{\alpha_j^A\} v_{I_j}}{n_T(\exp\{\alpha_j^A\} + v_{I_j})}\Big),\displaybreak[0]\\
\textstyle \gamma_j | \Omega \backslash \{\gamma_j\} & \textstyle \stackrel{ind.}{\sim} 
\begin{cases} 0, & I_j = 0, \\ N \big( \frac{\sigma_\gamma^2 \sum_i{(\lambda_{ij} - \mu_j^A)}}{v_1}, \frac{\sigma_\gamma^2 \exp\{\alpha_j^B\}}{v_1}\big), &I_j = 1, 
\end{cases} 
\end{align*}
for $j=1,\ldots,m$.
Further details and proofs can be found in Appendix \ref{postinferencedetails}.

Typically, we run the MCMC algorithm for 20,000 iterations, discarding the first 10,000 as burn-in steps and saving every 10th remaining step. In a setup with 10,000 genes and 2 samples in each group, one iteration takes around 0.13 seconds on a 2 GHz 64-bit processor with the BADER software. Computation times should change approximately linearly for other numbers of genes or samples, as the required number of computations for posterior inference is $\mathcal{O}(mn)$.

\section{Results}

\subsection{Inference on the Dispersion Parameter \label{sec:dispersion}}

To assess inference on dispersion parameters in a realistic case,
we used a large dataset published by \citet{Pickrell2010},
available from
\url{http://bowtie-bio.sourceforge.net/recount/} \citep{Frazee2011}.
Of the $69$ samples in the dataset, we took subsamples of $n \in \{2,5,10,20\}$,
corresponding to typical sample sizes in RNA-seq datasets. Genes
where the sum of the remaining counts for all samples were less or equal to 5
were dropped from the analysis. We assumed that the posterior medians of
$\mu_j$ and $\alpha_j$ from the full dataset were the ``true'' values, and
calculated the empirical coverage of central $80\%$ posterior credible
intervals for these parameters derived from the subsamples. For both parameters
and for all (sub-)sample sizes, the empirical coverage of the ``true'' values
was between about $70\%$ and $90\%$, indicating accurate
calibrations of the posterior distributions of the gene-specific dispersion
parameters.

\subsection{Determining Differential Expression \label{sec:DE}}


\begin{figure}[h]
\centering\includegraphics[width=0.65\textwidth]{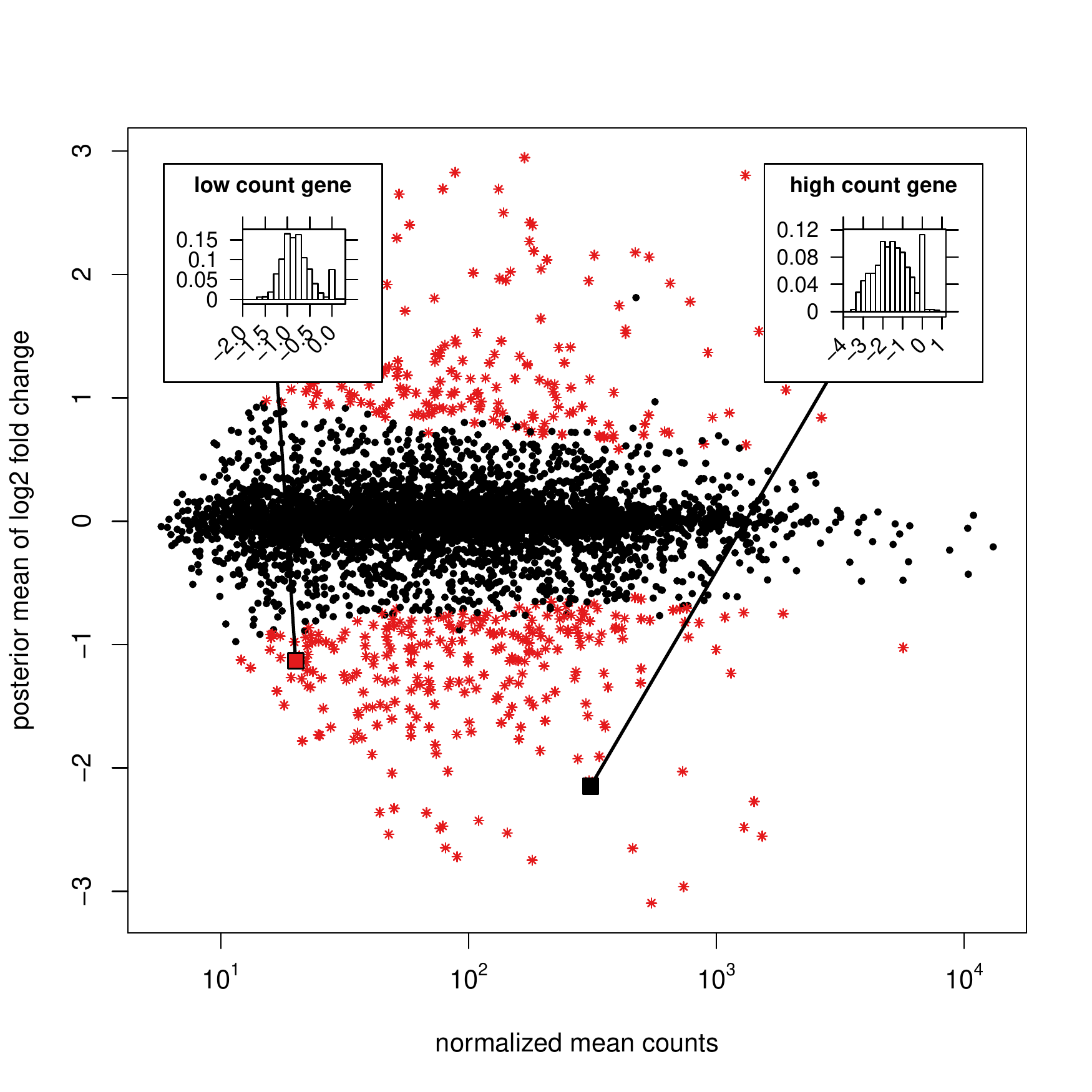}
\caption{Scatterplot of the posterior mean of the log fold change parameter from BADER against the normalized mean count for genes in the dataset from \citet{Katz2010}.
The red stars indicate genes with a posterior probability of DE higher than $0.9$. The plot inserts show the posterior distribution of the log fold change parameter of a gene found to be DE with a low mean count, and of a gene with a high absolute log fold change parameter but posterior probability of DE smaller than $0.9$.}
\label{scatterMouse}
\end{figure}

We ran BADER on a dataset from \citet{Katz2010}, where we dropped all genes
for which the sum of all counts were less or equal to 5. Figure \ref{scatterMouse} 
shows a scatterplot of the posterior means of the log fold change parameters against the normalized mean counts, with red stars indicating the genes with a posterior probability of DE higher than $0.9$. Interestingly, we find some
genes with high absolute value of the log fold change parameter but smaller
posterior probability of DE than $0.9$, due to the heterogeneous gene-specific posterior uncertainty allowed by our flexible model.

Because we do not know the ground truth, these results do not tell us whether BADER improves upon existing methods in terms of DE detection. Hence, we then compared methods on simulated datasets where we do know the underlying truth. We generated 100 datasets, each consisting of $n_A = n_B =2$ samples for $m=5000$ ``genes,'' divided into 250 sets of 20 genes each. Of these groups, $90\%$ were assumed to be not enriched and we chose a small probability of DE ($0.1$) for a gene in these groups. The remaining $10\%$ of groups were assumed to be enriched and consequently, genes in these sets were DE with a high probability of $0.75$.
The individual gene expression counts were simulated according to model \eqref{lognormalmodel} with hyperparameters $\psi_0 = -3$, $\tau = 0.8$, and $\sigma_\gamma = \log_2(1.5)$, which were chosen to be as realistic as possible. The mean log expression level for gene set $l$ was chosen as $\bar{\mu}^A_l \sim N(5,1)$ for $l=1,\ldots,250$, with individual gene log expression levels in set $l$ simulated as $\mu_j^A | \bar{\mu}^A_l \sim N(\bar{\mu}^A_l, 0.5^2)$. 
In this subsection, we will focus on determining DE for individual genes, and we will consider inference on enriched gene sets in Section \ref{sec:enrichment}.

\begin{figure}[h]
\centering\includegraphics[width=0.6\textwidth]{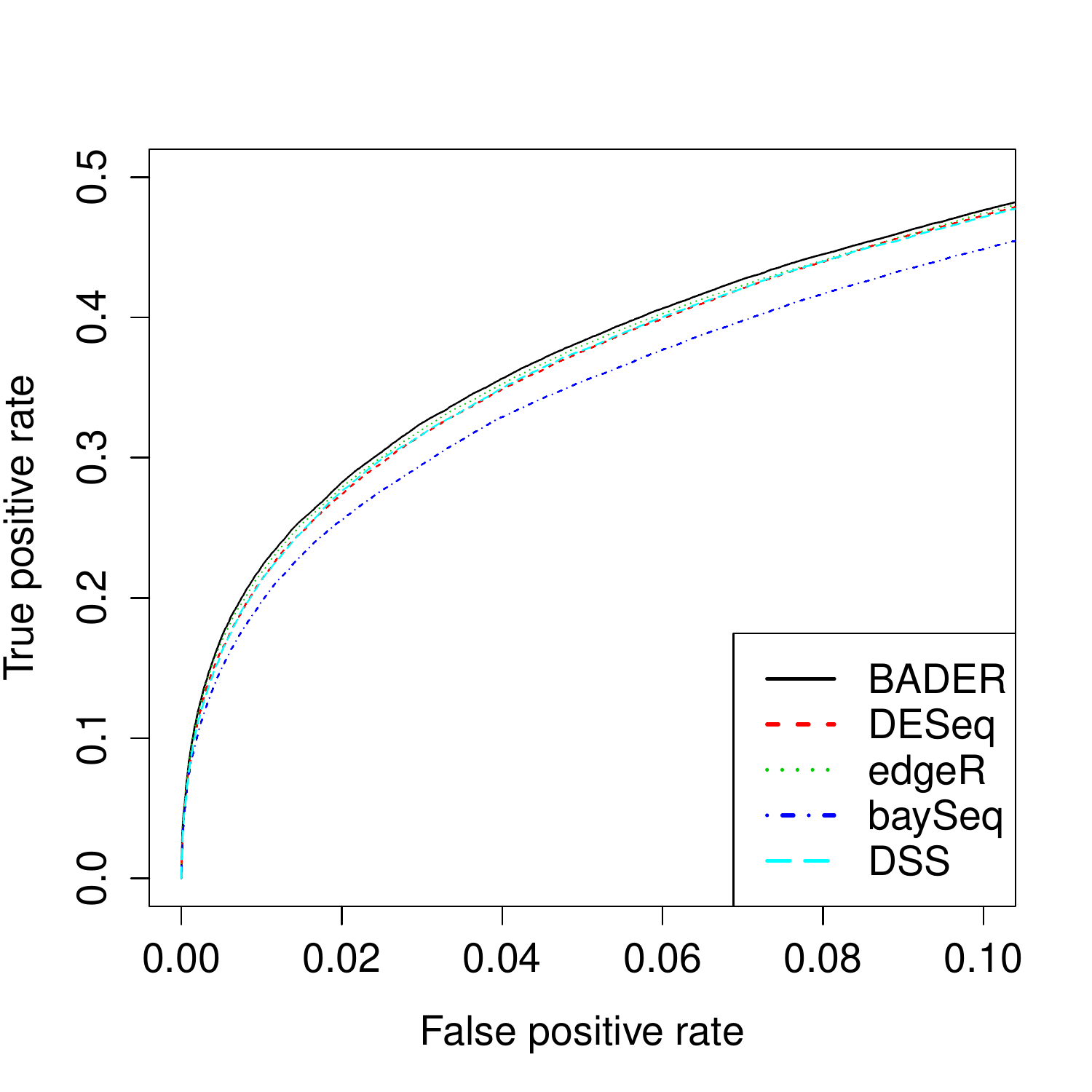}
\caption{DE-detection ROC curves for simulated data: Comparing BADER to four competitors}
\label{ROCfull}
\end{figure}

Using these 100 simulated datasets, we compared BADER (version 0.99.2) to edgeR \citep{robinson2010edger} (version 2.6.10), DESeq \citep{Anders2010} (version 1.12.1), baySeq
\citep{Hardcastle2010} (version 1.14.1), and DSS \citep{Wu2013} (version 1.4.0), applying the methods from the packages' vignettes. We compared results averaged over all 100 datasets using a receiver operating characteristic (ROC) curve, which plots the true positive rate versus the false positive rate when varying over a threshold parameter $q$
(Figure~\ref{ROCfull}). In the case of edgeR, DESeq, and DSS, we take all genes to be DE for which the p-value is lower than $q$. In the case of baySeq and BADER, we take genes to be DE if the corresponding posterior probability is larger than $1-q$. This analysis showed that BADER outperformed all other approaches in calling genes differentially expressed.

It is important to note again that, while BADER performs well in terms of DE detection, the output from our algorithm is much richer than just posterior DE
probabilities for individual genes. Among other things, we obtain the joint posterior distribution of the log-fold-change parameters $\gamma_j$ (see
Equation \eqref{foldchange}), which, for example, also contains the posterior distribution of the magnitude of the fold change for DE genes. In addition,
this output allows for further downstream analyses, an example of which is given in the next section.

\subsection{Gene Set Enrichment \label{sec:enrichment}}

Once we have obtained samples from the posterior distribution as described in Section \ref{sec:inference}, inference on sets or groups of genes (``gene set enrichment'') can proceed without much difficulty. For example, one could use the posterior distributions of DE indicators as an input to the algorithm of \citet{Bauer2010}, for which groups of genes are defined to be enriched if all genes in that group are DE. 

Here, for each set of genes $\mathcal{S} \subset \{1,\ldots,m\}$, we test the ``competitive null hypothesis'' \citep{Goeman2007} that the genes in $\mathcal{S}$ are at most as often differentially expressed as the remaining genes by considering the posterior probability of the event
\begin{equation}
\label{geneSetDE}
  \textstyle\sum_{j \in \mathcal{S}} I_j \, / \, | \mathcal{S} | \,> \, \sum_{j \in \mathcal{S}^c} I_j \, / \, | \mathcal{S}^c |,
\end{equation}
where the $I_j$ are the DE indicators introduced in \eqref{foldchange}.

A problem with gene set enrichment has been that for genes with low counts, the
power for DE detection is low. Hence, in testing for category enrichment,
categories or groups of genes with low counts are rarely determined to be
enriched. \citet{Young2010} and \citet{Mi2012} attempt to remedy this problem by accounting for differences in gene length. However, we believe that different gene lengths are not the only cause of bias in enrichment analysis, and it is necessary to account for differences in gene-specific DE uncertainty directly.

\begin{figure}[h]
\centering\includegraphics[width=0.6\textwidth]{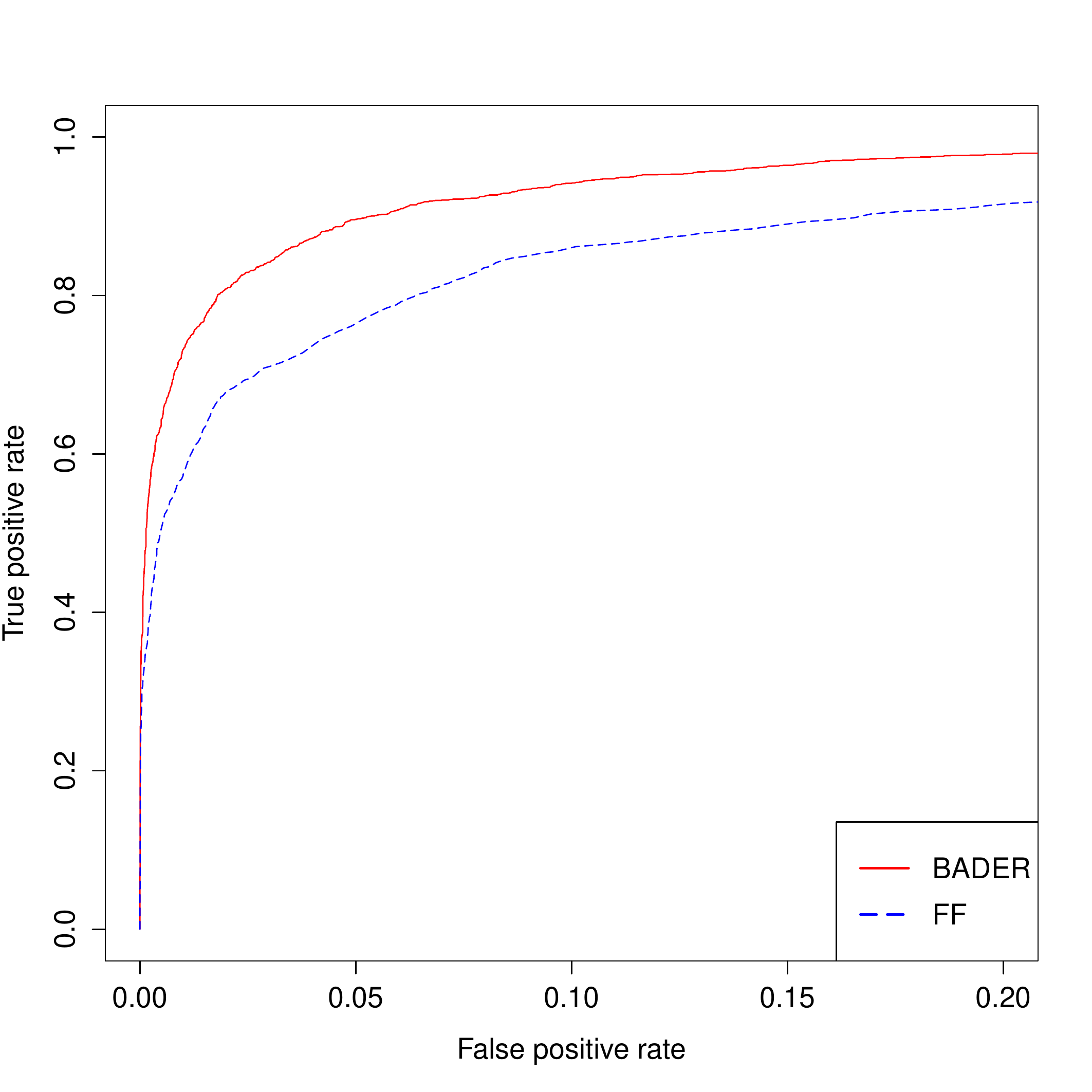}
\caption{ROC curves for detection of enriched gene sets: As determined by BADER in red, as determined by a combination of DESeq and Fisher' exact test (FF) in blue}
\label{geneSet_roc}
\end{figure}

We considered the same 100 simulated datasets as described in Section \ref{sec:DE}. For BADER, we obtained the posterior probability for a gene set to be enriched by estimating the posterior probability of \eqref{geneSetDE}. We compared our results to a frequentist approach (henceforth referred to as FF) that consisted of testing for DE for each gene using DESeq \citep{Anders2010} with a significance level of 0.05, followed by calculating the p-value of Fisher's exact test for over-representation of DE genes in the set under
consideration. Figure \ref{geneSet_roc} shows the ROC curves for the two methods. BADER considerably outperforms the FF approach. The difference in performance is especially large when focusing on gene sets with low mean expression level (e.g., $\bar{\mu}^A_l < 3$; figure not shown).

\section{Discussion}

We developed a fully Bayesian model for the analysis of RNA-seq
count data (BADER). We described proper MCMC-based posterior inference, taking into account all uncertainty without shortcuts.
BADER is distributed open-source on Bioconductor as an \texttt{R} package with an efficient \texttt{C++}
back-end.

We demonstrated the value of posterior samples of our log-fold-change parameters with a point mass at zero. Their use in
DE detection is highly competitive and natural, avoiding explicit
adjustments for multiple testing. Moreover, their integration with
downstream analyses is not only conceptually easier, but also suffers
less from biases than many frequentist approaches. We demonstrated this point
by using BADER posterior samples for gene set
enrichment analysis, showing more power to detect enriched gene
sets.


\section*{Acknowledgments}

We thank Daniel Bader, Jens Preussner, and several anonymous reviewers for helpful feedback on the
manuscript. J.G.\ is supported by the Bavarian Research Center for Molecular Biosystems.


\newpage
\appendix

\small

\section{Some details on posterior inference \label{postinferencedetails}}

Posterior inference is summarized in Section 2.2 of the main article. Here we derive the updating steps that allow joint sampling from the distribution $[\gamma_j, \mu_j^A, I_j | \cdot ]$. This is achieved by integrating out both $\mu_j^A$ and $\gamma_j$ when updating $I_j$, and integrating out $\gamma_j$ when updating $\mu_j^A$.

Again, let $\Omega$ denote the set of all parameters, $\overline{\lambda_j^T} \colonequals \sum_i \lambda_{ij}^T/n_T$, and $P_{\mu_j^A}$ and $P_{\gamma_j}$ are the measures corresponding to the prior distributions of $\mu_j^A$ and $\gamma_j$, respectively. We begin by proving the following proposition:

\begin{proposition}\label{normalprop}
\[
	N(x|\mu_1,\sigma_1^2)	N ( x|\mu_2,\sigma_2^2)  = N(\mu_1 | \mu_2 , \sigma_1^2 + \sigma_2^2) N \left( x \left| \frac{\mu_1 \sigma_2^2 + \mu_2 \sigma_1^2}{\sigma_1^2 + \sigma_2^2},\frac{\sigma_1^2 \sigma_2^2}{\sigma_1^2 + \sigma_2^2} \right) \right.
\]
\begin{proof}
    \begin{align*}
N(x|\mu_1,&\sigma_1^2)N ( x|\mu_2,\sigma_2^2) \\
    &= \frac{1}{\sqrt{2 \pi \sigma_1^2}} \exp\left(\frac{(x-\mu_1)^2}{2 \sigma_1^2}\right) \frac{1}{\sqrt{2 \pi \sigma_2^2}} \exp \left(\frac{(x-\mu_2)^2}{2 \sigma_2^2}\right)  \\
    &=  \frac{1}{2 \pi \sigma_1 \sigma_2} \exp\left(\frac{(\sigma_1^2 + \sigma_2^2)x^2 - 2(\mu_1 \sigma_2^2 + \mu_2 \sigma_1^2)x + \mu_1^2 \sigma_2^2 + \mu_2^2 \sigma_1^2}{2 \sigma_1^2 \sigma_2^2}\right) \displaybreak[0]\\
    &=  \frac{1}{2 \pi \sigma_1 \sigma_2} \exp\left(\frac{\left(x-\frac{\mu_1 \sigma_2^2 + \mu_2 \sigma_1^2}{\sigma_1^2 + \sigma_2^2}\right)^2}{2 \frac{\sigma_1^2 \sigma_2^2}{\sigma_1^2 + \sigma_2^2}}\right) \\
    &\quad \quad \quad \times \exp\left( \frac{\sigma_1^2 + \sigma_2^2}{2 \sigma_1^2 \sigma_2^2} \left(\frac{\mu_1^2 \sigma_2^2 + \mu_2^2 \sigma_1^2}{\sigma_1^2 + \sigma_2^2} - \left( \frac{\mu_1 \sigma_2^2 + \mu_2 \sigma_1^2}{\sigma_1^2 + \sigma_2^2}\right)^2\right) \right) \\
    &=  N \left( x \left| \frac{\mu_1 \sigma_2^2 + \mu_2 \sigma_1^2}{\sigma_1^2 + \sigma_2^2},\frac{\sigma_1^2 \sigma_2^2}{\sigma_1^2 + \sigma_2^2} \right) \right. \\
    &\quad \quad \quad \times \frac{1}{\sqrt{2 \pi} \sqrt{\sigma_1^2 + \sigma_2^2}} \exp\left(\frac{\mu_1^2 \sigma_1^2 \sigma_2^2 + 2 \mu_1 \mu_2 \sigma_1^2 \sigma_2^2 + \mu_2^2 \sigma_1^2 \sigma_2^2}{2 \sigma_1^2 \sigma_2^2(\sigma_1^2 + \sigma_2^2)}\right) \\
    &= N \left( x \left| \frac{\mu_1 \sigma_2^2 + \mu_2 \sigma_1^2}{\sigma_1^2 + \sigma_2^2},\frac{\sigma_1^2 \sigma_2^2}{\sigma_1^2 + \sigma_2^2} \right) \right. N(\mu_1 | \mu_2 , \sigma_1^2 + \sigma_2^2) .
\end{align*}
\end{proof}
\end{proposition}

\noindent Then we have
\begin{align*}
[I_j | \Omega \backslash \{\mu_j^A, \gamma_j, I_j\} ] &= \iint [I_j | \Omega \backslash I_j] \,dP_{\mu_j^A} \,dP_{\gamma_j} \\
& \propto \iint \pi^{I_j} (1-\pi)^{1-I_j} \prod_i N(\lambda_{ij}^A | \mu_j^A,e^{\alpha_j^A}) N(\lambda_{ij}^B | \mu_j^A + \gamma_j,e^{\alpha_j^B}) \,d\mu_j^A \,dP_{\gamma_j} \\
& \propto \int \pi^{I_j} (1-\pi)^{1-I_j} N\Big(\mu_j^A \Big| \overline{\lambda_{j}^A},\frac{exp(\alpha_j^A)}{n_T}\Big) N\Big(\mu_j^A \Big| \overline{\lambda_j^B} - \gamma_j,\frac{exp(\alpha_j^B)}{n_T}\Big) dP_{\gamma_j} \\
& \stackrel{Prop.~\ref{normalprop}}{\propto} \int \pi^{I_j} (1-\pi)^{1-I_j} N\Big(\overline{\lambda_j^A} \Big| \overline{\lambda_j^B} - \gamma_j, \frac{exp({\alpha_j^A})+ exp({\alpha_j^B})}{n_T}\Big) \,dP_{\gamma_j}.
\end{align*}
Applying Proposition~\ref{normalprop} again yields
\begin{align*}
P( I_j = 1 | \Omega \backslash \{\mu_j^A, \gamma_j, I_j\} ) &\propto \pi N(\overline{\lambda_j^A} | \overline{\lambda_j^B}, (e^{\alpha_j^A} + e^{\alpha_j^B})/n_T + \sigma_\gamma^2), \\
P( I_j = 0 | \Omega \backslash \{\mu_j^A, \gamma_j, I_j\} ) &\propto (1-\pi) N(\overline{\lambda_j^A} | \overline{\lambda_j^B}, (e^{\alpha_j^A} + e^{\alpha_j^B})/n_T),
\end{align*}
which is a Bernoulli distribution with parameter $\frac{\pi N_1}{ \pi N_1 + (1-\pi) N_0 }$, where
\begin{align*}
N_0 \colonequals ~&N(\overline{\lambda_j^A} | \overline{\lambda_j^B}, (e^{\alpha_j^A} + e^{\alpha_j^B})/n_T), \\
N_1 \colonequals ~&N(\overline{\lambda_j^A} | \overline{\lambda_j^B}, (e^{\alpha_j^A} + e^{\alpha_j^B})/n_T + \sigma_\gamma^2).
\end{align*} 
To obtain $[\mu_j^A | \Omega \backslash \{\mu_j^A, \gamma_j \}]$, note that 
\begin{align*}
[\mu_j^A | \Omega \backslash \mu_j^A ] &\propto \prod_{i,T} N(\lambda_{ij}^T | \mu_j^A + \gamma_j I(T=B),e^{\alpha_j^T})
\intertext{and}
\gamma_j &\sim \left\{ \begin{array}{cl} 0, & I_j = 0, \\ N(0,\sigma_\gamma^2), & I_j = 1, \end{array} \right.
\end{align*}
and so, applying Proposition~\ref{normalprop} again, we have
\begin{align*}
[\mu_j^A | \Omega \backslash \{\mu_j^A, \gamma_j \}] &\propto \left\{ \begin{array}{lc}
C_0 N\left( \mu_j^A \left| \frac{\overline{\lambda_j^A} e^{\alpha_j^B} + \overline{\lambda_j^B} e^{\alpha_j^A}}{e^{\alpha_j^A} + e^{\alpha_j^B}},\frac{e^{\alpha_j^A} e^{\alpha_j^B}}{n_T (e^{\alpha_j^A} + e^{\alpha_j^B})}\right) \right. , & I_j = 0, \\
C_1 N \left(\mu_j^A \left| \frac{\overline{\lambda_j^A} ( e^{\alpha_j^B} + n_T \sigma_\gamma^2) + \overline{\lambda_j^B} e^{\alpha_j^A}}{e^{\alpha_j^A} + e^{\alpha_j^B} + n_T \sigma_\gamma^2},\frac{e^{\alpha_j^A}( e^{\alpha_j^B} + n_T \sigma_\gamma^2)}{n_T (e^{\alpha_j^A} + e^{\alpha_j^B} + n_T \sigma_\gamma^2)}\right) \right. , & I_j = 1,
\end{array} \right.
\end{align*}
where
\begin{align*}
C_0 &\colonequals \pi N(\overline{\lambda_j^A} | \overline{\lambda_j^B}, (e^{\alpha_j^A} + e^{\alpha_j^B})/n_T), \\
C_1 &\colonequals (1-\pi) N(\overline{\lambda_j^A}| \overline{\lambda_j^B},(e^{\alpha_j^A} + e^{\alpha_j^B})/n_T + \sigma_\gamma^2) .
\end{align*}
As $C_0$ and $C_1$ do not contain $\mu_j^A$, we can write this as,
\begin{align*}
\mu_j^A |  \Omega \backslash \{\mu_j^A, \gamma_j \} &\stackrel{ind.}{\sim} \left\{ \begin{array}{lc}
N\left( \frac{\overline{\lambda_j^A} e^{\alpha_j^B} + \overline{\lambda_j^B} e^{\alpha_j^A}}{e^{\alpha_j^A} + e^{\alpha_j^B}},\frac{e^{\alpha_j^A} e^{\alpha_j^B}}{n_T (e^{\alpha_j^A} + e^{\alpha_j^B})}\right) , & I_j = 0, \\
N \left( \frac{\overline{\lambda_j^A} ( e^{\alpha_j^B} + n_T \sigma_\gamma^2) + \overline{\lambda_j^B} e^{\alpha_j^A}}{e^{\alpha_j^A} + e^{\alpha_j^B} + n_T \sigma_\gamma^2},\frac{e^{\alpha_j^A}( e^{\alpha_j^B} + n_T \sigma_\gamma^2)}{n_T (e^{\alpha_j^A} + e^{\alpha_j^B} + n_T \sigma_\gamma^2)}\right) , & I_j = 1 .
\end{array} \right.
\end{align*}
Finally, we have
\begin{equation*}
[ \gamma_j | \Omega \backslash \{\gamma_j\} ] \propto \left\{ \begin{array}{cl} 0 , & I_j = 0, \\ \left( \prod_i N(\lambda_{ij}^B | \mu_j^A + \gamma_j, e^{\alpha_j^B})\right) N(\gamma_j | 0, \sigma_{\gamma}^2), & I_j = 1, \end{array} \right.
\end{equation*}
which can be shown to be a mixture mixture of a point mass at zero and a normal distribution:
\begin{equation*}
\gamma_j | \Omega \backslash \{\gamma_j\} \stackrel{ind.}{\sim} \left\{\begin{array}{cl} 0, & I_j = 0, \\ 
N \left( \frac{\sigma_\gamma^2 \sum_i{(\lambda_{ij}^B - \mu_j^A)}}{n \sigma_\gamma^2 + e^{\alpha_j^B}}, \frac{\sigma_\gamma^2 e^{\alpha_j^B}}{e^{\alpha_j^B} + n_T \sigma_\gamma^2}\right), &I_j = 1. \end{array} \right.
\end{equation*}


\section{Comparison between our Poisson-lognormal data model and a negative-binomial data model \label{app:comparison}}

\subsection{Analytic comparison\label{app:analyticcomparison}}

For the analysis of RNAseq counts, the data model is often \citep[e.g.,][]{Robinson2008,Anders2010}
assumed to be of the form,
\begin{equation}
\label{marginaldatamodel}
\begin{split}
k_{ij}^T | \mu_j^T,\alpha_j^T \stackrel{ind.}{\sim}
	NegBinom( s_i^T e^{\mu_j^T} , e^{\alpha_j^T} );\quad
	i=1,\ldots,n_T;~j=1,\ldots,m;~T = A,B, 
\end{split}
\end{equation}
instead of the two-stage Poisson-lognormal described in Section 2.1 of the main article. However, the two models are closely related. 
Model \eqref{marginaldatamodel} can also be written in two stages \citep[e.g.,][]{Anders2010}:
\begin{equation}
\label{datamodel1}
\begin{split}
   k_{ij}^T | \lambda_{ij}^T & \stackrel{ind.}{\sim} Poi( s_i^T e^{\lambda_{ij}^T}); \quad i=1,\ldots,n_T; \, j=1,\ldots,m; \, T = A,B,\\
 e^{\lambda_{ij}^T} | \mu_j^T, \alpha_j^T & \stackrel{ind.}{\sim} Gam( e^{\mu_j^T}, e^{\alpha_j^T}  ); \quad i=1,\ldots,n_T; \, j=1,\ldots,m; \, T = A,B,
\end{split}
\end{equation}
where $Gam(a,b)$ denotes a gamma distribution with mean $a$, squared coefficient of variation $b$,
and variance $a^2b$. Clearly, like the Poisson-lognormal model in Section 2.1 of the main article, this negative-binomial data model
is also a type of overdispersed Poisson model with overdispersion parameter $\alpha_j^T$. The difference between the two data models is the type of distribution describing the overdispersion (lognormal versus gamma). These two distributions are fairly similar. In fact, it has been shown that inference using the lognormal distribution
will typically produce the same conclusions as inference using the gamma distribution (\citealp{Atkinson1982}; \citealp{McCullagh1989}, pp.\ 286/293).

\subsection{Numerical Comparison\label{app:numericalcomparison}}

In Section \ref{postinferencedetails} we have seen that the Poisson-lognormal data model allows closed-form updating of the heavily dependent parameters $\gamma_j$, $\mu_j^A$, and $I_j$ in the MCMC sampler. When implementing an MCMC sampler for the negative-binomial model \eqref{marginaldatamodel}, this is not possible, and we need to resort to Metropolis-Hastings (MH) updates for these three parameters. We update the parameters jointly, using an extension of the adaptive MH sampler of \citet{Haario2001}, to account for that fact that $I_j$ follows a discrete distribution: After an initial burn-in phase, the proposed values for the three parameters at the $(l+1)$-th MCMC iteration are drawn from,
\begin{align*}
I_j^{(l*)} &\sim B( \overline{I_{l-1}}), \\
\left( \begin{array}{c} \gamma_j^{(l*)} \\ \mu_j^{A (l*)} \end{array} \right) \Big| \,I_j^{(l*)} &\sim 
\left\{ \begin{array}{cl} (\delta_0, N(\mu_j^{A (l_0)},\Sigma_0^{(l)}))', & I_j^{(l*)} = 0, \\ 
N((\gamma_j^{(l_1)},\mu_j^{A (l_1)})',\Sigma_1^{(l)}), & I_j^{(l*)} = 1, \end{array} \right.
\end{align*}
where the superscript ${}^{(l)}$ denotes parameter values at the previous iteration $l$, $l_0 = l_0(l)$ and $l_1 = l_1(l)$ denote the maximum of all $q \leq l$ for which $I_j^{(q)} = 0$ and $I_j^{(q)} = 1$, respectively, and $\Sigma_0^{(l)}$ and $\Sigma_1^{(l)}$ are calculated following \citet{Haario2001}.

To compare the two data models numerically regarding mixing, speed, and accuracy, we simulated data separately from each model using the following setup: $m=1000$, $n \in \{2,5,10,20\}$, $\pi_0 = 0.5$, $\psi_0 = -3$, $\sigma_\gamma = 0.8$, and $\tau = 0.8$.

In initial tests, the negative binomial model did not converge. With the noninformative prior on $\tau$ (see the end of Section 2.1 in the main article), this parameter drifted off to $+\infty$ as some $\alpha_j^T$ tended to $-\infty$ (indicating no overdispersion). Hence, we had to leave $\tau$ fixed for the negative-binomial model. To ensure comparability, we did so for the lognormal model in this simulation study as well.

We simulated a dataset from this setup for each of the models, and fitted the models for each dataset using 5,000 burn-in steps and 10,000 subsequent MCMC iterations, of which we saved every 10th step for inference. As a measure of yield per time unit, we took the effective sample size (ESS) per minute of $\mu_j^T$ and $\alpha_j^T$. The ESS estimates the number of independent samples that would have the equivalent amount of information as the actual MCMC sample (in our case, of size 1,000), by correcting for autocorrelation \citep[we use the implementation in the R package CODA; see][]{Plummer2006}. As measure of accuracy, we used the average (over all genes) continuous ranked probability score \citep[CRPS; see][]{Gneiting2007}.

Results of the simulation study are shown in Table~\ref{logVsNeg}. The yield of effective samples of $\alpha_j^T$ and $\mu_j^T$ per minute is much higher for the lognormal than for the negative-binomial model. Even without taking run time into account, the posterior distributions from the former model yield lower average CRPS, indicating higher accuracy. (But since the parameters in the two models are not exactly the same, the CRPS might not be directly comparable.)

\begin{table}
\centering
\begin{minipage}[t]{\textwidth}
\centering
\begin{tabular}{c|c|c|c|c|c|c}
& model & time & ESS/min ($\alpha_j^T) $ & ESS/min ($\mu_j^T$) & CRPS ($\alpha_j^T)$ & CRPS ($\mu_j^T)$ \\
\hline
\hline  
\multirow{2}{*}{$n=2$} & lognormal & 5.4 & 166.9 & 143.9 & 0.412 & 0.114 \\
\cline{2-7} & neg-binom & 10.7 & 2.3 & 86.6 & 0.892 & 0.116 \\
\hline
\hline
\multirow{2}{*}{$n=5$} & lognormal & 8.0 & 99.0 & 96.4 & 0.344 & 0.074 \\
\cline{2-7} & neg-binom & 14.5 & 11.4 & 63.6 & 0.533 & 0.081 \\
\hline
\hline
\multirow{2}{*}{$n=10$} & lognormal & 12.1 & 57.8 & 62.6 & 0.304 & 0.051 \\
\cline{2-7} & neg-binom &  20.3 & 16.2 & 38.2 & 0.450 & 0.053\\
\hline
\hline
\multirow{2}{*}{$n=20$} & lognormal & 20.8 & 30.3 & 36.2 & 0.264 & 0.037 \\
\cline{2-7} & neg-binom & 33.6 & 6.2 & 14.2 & 0.475 & 0.038 \\
\hline
\end{tabular}
\caption{Comparison of the lognormal data model and the negative-binomial
model using simulated data. Time is shown in minutes, effective sample size (ESS) per minute and CRPS
are averaged over all genes.}
\label{logVsNeg}
\end{minipage}
\end{table}

We also tried applying the negative-binomial model to the dataset of \citet{Pickrell2010}, holding $\tau$ fixed at a point estimate from a large dataset. However, we encountered even more severe mixing problems than for the simulated data. The traceplots of the dispersion parameters, $\alpha_j^T$, did not indicate convergence even after 30,000 MCMC iterations.

\newpage
\bibliographystyle{apalike}
\bibliography{template_bib}
\end{document}